\documentstyle[12pt]{article}
\newsavebox{\trifig}

\newcommand{\refjl}[4]{{#1 }{\em #2 }{\bf #3 }{#4}}
\newcommand{\refbk}[3]{{#1 }{\em #2 }{#3}}

\def\JPA{{J. Phys. A:Math. Gen.}}
\def\JPC{{J. Phys. C:Solid State Phys.}}

\def\PRB{{Phys. Rev. B}}

\begin{document}
\setcounter{page}{0}
\thispagestyle{empty}

\title{Series expansions of the percolation probability
on the directed triangular lattice.}

\author{
Iwan Jensen and Anthony J. Guttmann \\
Department of Mathematics \\
University of Melbourne \\
Parkville, Victoria 3052 \\
Australia. \\
e-mail: iwan, tonyg@maths.mu.oz.au
}

\maketitle

\thispagestyle{empty}

\begin{abstract}
We have derived long series expansions of the percolation
probability for site, bond and site-bond percolation on the
directed triangular lattice. For the bond problem we have
extended the series from order 12 to 51 and for the site
problem from order 12 to 35. For the site-bond problem, which
has not been studied before, we have derived the series to
order 32. Our estimates of the critical exponent $\beta$ are
in full agreement with results for similar problems on the
square lattice, confirming expectations of universality. For
the critical probability and exponent we find in the site
case: $q_c = 0.4043528 \pm 0.0000010$ and
$\beta = 0.27645 \pm 0.00010$; in the bond case:
$q_c = 0.52198\pm 0.00001$ and $\beta = 0.2769\pm 0.0010$;
and in the site-bond case:
$q_c = 0.264173 \pm 0.000003$ and $\beta = 0.2766 \pm 0.0003$.
In addition we have obtained accurate estimates for the
critical amplitudes. In all cases we find that the leading
correction to scaling term is analytic, i.e., the confluent
exponent $\Delta = 1$.
\end{abstract}

\vspace{20mm}


\newpage

\section{Introduction}

In an earlier paper (Jensen and Guttmann 1995) we reported
on the derivation and analysis of long series for the
percolation probability of site and bond percolation on
the directed square and hexagonal lattices. In this paper
we extend this work to site, bond and site-bond percolation
on the directed triangular lattice. We refer to our earlier
paper for a more general introduction to directed percolation
and its role in the modelling of physical systems.
In directed {\em site}
percolation each site is either present (with probability $p$)
or absent (with probability $q=1-p$) independent of all other
sites on the lattice.  Similarly for {\em bond} percolation each
bond is absent or present independently of other bonds. Finally
in {\em site-bond} percolation both sites and bonds may be absent
or present with equal probability, but again with no dependency
on any other sites or bonds. Two sites in the various models
are connected if one can
find a path, respecting the directions indicated in
Figure~\ref{fig-oriented}, through occupied sites, bonds or sites
{\em and} bonds, respectively,
from one to the other. When $p$ is smaller than a critical value
$p_c$ all clusters of connected sites remain finite, while for
$p \geq p_c$ there is an infinite cluster spanning the lattice
in the preferred direction. The order parameter of the system
is the percolation probability $P(p)$ that a given site belongs
to the infinite cluster. This quantitiy is strictly zero when
$p < p_c$ and changes continuously at $p_c$. For $p > p_c$ the
behaviour of $P(p)$ in the vicinity of $p_c$ may be described
by a critical exponent $\beta$,

\begin{equation}
  P(p) \propto (p-p_c)^{\beta}, \;\;\;\;  p \rightarrow p_c^+.
\end{equation}

The bond problem was originally studied by Blease (1977) who
calculated a series to 12th order. For the site problem
De'Bell and Essam (1983) derived the series to 12th order.
The site-bond problem has, at least to our knowledge, never
been studied before. Our main motivation for doing so
in this paper is to obtain further independent estimates of the
critical exponent $\beta$.
Using the finite-lattice method and the
extrapolation technique of Baxter and Guttmann (1988) we have
extended the series for the bond problem to order 51, for the
site problem to order 35 and derived the series for the
site-bond problem to order 32. The site and bond problems have
also been studied by Essam {\em et al.} (1986,1988), who
derived series expansions for moments of the pair connectedness.

\section{The finite-lattice method}

\begin{figure}[tb]
\begin{picture}(420,220)
\savebox{\trifig}(60,35){
\put(30,35){\circle*{3}}
\put(10,23){\vector(-3,-2){2}}
\put(50,23){\vector(3,-2){2}}
\put(30,13){\vector(0,-1){2}}
\put(30,35){\line(-5,-3){30}}
\put(30,35){\line(0,-1){35}}
\put(30,35){\line(5,-3){30}}}

\multiput(150,25)(0,35){5}{\usebox{\trifig}}
\multiput(120.5,42.5)(0,35){4}{\usebox{\trifig}}
\multiput(179.5,42.5)(0,35){4}{\usebox{\trifig}}
\multiput(91,60)(0,35){3}{\usebox{\trifig}}
\multiput(209,60)(0,35){3}{\usebox{\trifig}}
\multiput(61.5,77.5)(0,35){2}{\usebox{\trifig}}
\multiput(238.5,77.5)(0,35){2}{\usebox{\trifig}}
\multiput(32,95)(0,35){1}{\usebox{\trifig}}
\multiput(268,95)(0,35){1}{\usebox{\trifig}}
\multiput(210,25)(29.5,17.5){6}{\circle*{3}}
\multiput(210,25)(-29.5,17.5){6}{\circle*{3}}

\put(206,210){O}

\multiput(210,25)(2,1.1864){80}{\circle*{0.1}}
\multiput(210,25)(-2,1.1864){80}{\circle*{0.1}}
\multiput(328,130)(2,1.1864){6}{\circle*{0.1}}
\multiput(92,130)(-2,1.1864){6}{\circle*{0.1}}
\multiput(298.5,147.5)(2,1.1864){6}{\circle*{0.1}}
\multiput(121.5,147.5)(-2,1.1864){6}{\circle*{0.1}}
\multiput(269,165)(2,1.1864){6}{\circle*{0.1}}
\multiput(150,165)(-2,1.1864){6}{\circle*{0.1}}
\multiput(239.5,182.5)(2,1.1864){6}{\circle*{0.1}}
\multiput(180.5,182.5)(-2,1.1864){6}{\circle*{0.1}}
\multiput(210,200)(2,1.1864){6}{\circle*{0.1}}
\multiput(210,200)(-2,1.1864){6}{\circle*{0.1}}
\put(15,117){$N=6$}
\put(378,117){$N=6$}
\put(44.5,135){$N=5$}
\put(343.5,135){$N=5$}
\put(74,152.5){$N=4$}
\put(314,152.5){$N=4$}
\put(103.5,170){$N=3$}
\put(284.5,170){$N=3$}
\put(133,187.5){$N=2$}
\put(253,187.5){$N=2$}
\put(162.5,205){$N=1$}
\put(232.5,205){$N=1$}

\put(215,92){\small $\sigma_t$}
\put(215,57){\small $\sigma_c$}
\put(165,75){\small $\sigma_l$}
\put(245,75){\small $\sigma_r$}
\end{picture}
\caption{ \label{fig-oriented} \sf
The directed triangular lattice with orientation given by the
arrows. The rows are labelled according to the text.}
\end{figure}
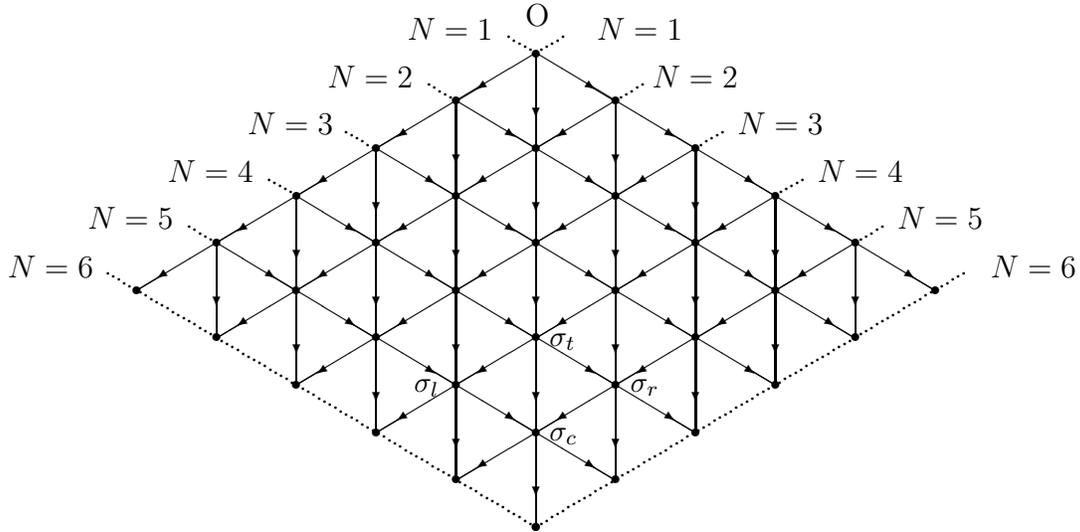

We wish to derive a series expansion for the percolation
probability on the directed triangular lattice oriented as in
Figure~\ref{fig-oriented}. In this figure we have numbered
the various levels or rows of the lattice according to which
sites can be reached by a path of minimum length $N-1$ starting
at the origin O. In other words all sites in the $N$th row
can be reached in $N-1$ steps but not in $N-2$ steps. Note that
a path going through a given site can only reach the part of
the lattice shown in Figure~\ref{fig-oriented} below the
origin O. This suggests that one should look at the following
finite-lattice approximation to $P(q)$, namely the probability
$P_N(q)$ that the origin is connected to at least one site in
the $N$th row. Since we are in the high density region we have
chosen to use the expansion parameter $q$ rather than $p$.
$P_N(q)$ is a polynomial with integer coefficients and a maximal
order determined by the total number of sites and/or bonds on
the finite lattice.

By the method used for the square lattice problems
(Bousquet-M\'{e}lou 1995) one can prove, {\em mutatis mutandis},
that the polynomials $P_N(q)$ converge to $P(q)$. Indeed we may
consider $P(q) = \lim_{N \rightarrow \infty} P_{N}(q)$ to be a
more precise definition of the percolation probability. More
importantly, however, from a series expansion point of view,
for the site and site-bond problems the first $N+1$ coefficients
of the polynomials $P_N(q)$ are identical to those of $P(q)$.
In the case of bond percolation the agreement extends through
the first $2N+1$ coefficients.

\subsection{Specification of the models}

To calculate the finite-lattice percolation probability $P_N(q)$
we associate a state $\sigma_j$ with each site, such that
$\sigma_j = 1$ if site $j$ is connected to the $N$th row
and $\sigma_j = -1$ otherwise. We shall often write
$+/-$ for simplicity.  Let $l$, $c$ and $r$ denote the sites
connected to a site $t$ from the row above,
as in Figure~\ref{fig-oriented}. We then define the weight
function $W(\sigma_t|\sigma_l, \sigma_c, \sigma_r)$ as the
probability that the top site $t$ is in state $\sigma_t$, given
that the lower sites $l$, $c$ and $r$ are in states
$\sigma_l$, $\sigma_c$ and $\sigma_r$, respectively. As for the
square lattice (Bidaux and Forgacs 1984, Baxter and Guttmann 1988)
we then have

\begin{equation}
   P_N(q) = \sum_{\{\sigma\}}\prod_t
            W(\sigma_t|\sigma_l,\sigma_c,\sigma_r),
   \label{eq:pnq}
\end{equation}

where the product is over all sites $t$ of the lattice above the
$N$th row. The sum is over all values $\pm 1$ of each $\sigma_t$,
other than the topmost spin $\sigma_1$ which always takes the
value +1. The spins in the $N$th row are fixed at +1, and
$P_N(q)$ is calculated as the sum over all possible configurations
of the probability of each individual configuration.

The weight functions $W$ are calculated as follows. Obviously,
$W(-|\sigma_l,\sigma_c,\sigma_r) =
1-W(+|\sigma_l,\sigma_c,\sigma_r)$.
The remaining weights are easily calculated by considering the
possible arrangements of states and sites and/or bonds.
$W(+|-,-,-)=0$ because the top site is connected to the $N$th
row if and only if at least one of its neighbours below
is connected to the $N$th row. All the remaining weights for
the {\em site} problem equal $1-q$ because the top-site has to
be occupied in order to be connected to the $N$th row.
Let us next look at the remaining {\em bond} weights.
$W^B(+|+,+,+) = 1-q^3$ because the only bond configuration
{\em not} allowed is all three bonds absent, which has
probability $q^3$. $W^B(+|+,+,-) = W^B(+|+,-,+) =
W^B(+|-,+,+) = 1-q^2$ because the bond to the $-$ state can
be either present or absent (probability 1) while among the
remaining bonds only the configuration with both bonds absent
(probability $q^2$) is forbidden. Finally, $W^B(+|+,-,-) =
W^B(+|-,+,-)= W^B(+|-,-,+) = 1-q$ because the bond to the +
state has to be present, which happens with probability $p=1-q$,
while the other bonds can be either present or absent. For the
{\em site-bond} problem we find that
$W^{SB}(+|\sigma_l,\sigma_c,\sigma_r) =
(1-q)W^B(+|\sigma_l,\sigma_c,\sigma_r)$
because if the top state is +1 the top site has to be present.

\subsection{Series expansion algorithm}

Computer algorithms for the calculation of $P_N(q)$ are readily
found. These are basically implementations of the transfer
matrix technique. The general features of these algorithms
were described in our earlier paper (Jensen and Guttmann 1995), to
which we refer for further details. The sum over configurations
is performed by moving a boundary line through the lattice.
For each configuration along the boundary line one maintains a
(truncated) polynomial which equals the sum of the product of
weights over all possible states on the side of the boundary
already traversed. The boundary is moved through the lattice
one site at a time. The calculation of $P_N (q)$ by this method
is limited by memory, since one needs storage for $2^N$ boundary
configurations. However, as was the case with the square lattice,
this problem can be circumvented by introducing a cut into the lattice.
For each fixed configuration of states on this cut one evaluates
the lattice sum $P_N^C(q)$ and gets $P_N (q) = \sum_C P_N^C(q)$ as
the sum over all configurations of the cut. By placing the cut
appropriately, the growth in memory requirements can be reduced
to $2^{N/2}$.

\begin{figure}[tb]
\begin{picture}(420,260)

\multiput(240,42)(30,18){6}{\line(-5,3){180}}
\multiput(180,42)(-30,18){6}{\line(5,3){180}}
\multiput(210,24)(-30,18){6}
{\multiput(0,0)(30,18){6}{\line(0,1){36}}}

\multiput(60,132)(30,-18){6}
{\multiput(0,0)(30,18){6}{\vector(0,-1){2}}}
\multiput(45,141)(30,-18){6}
{\multiput(0,0)(30,18){6}{\vector(-3,-2){2}}}
\multiput(75,141)(30,-18){6}
{\multiput(0,0)(30,18){6}{\vector(3,-2){2}}}

\put(206,252){O}
\put(140,82){C}
\put(213,115){E}
\put(50,150){L}
\put(108,185){L'}
\put(167,222){L"}
\put(306,185){R}
\put(273,152){S}
\put(332,117){B}

\multiput(150,96)(30,18){3}{\circle*{6}}
\multiput(240,150)(30,18){2}{\circle{6}}
\multiput(240,186)(30,18){1}{\circle{6}}
\multiput(210,204)(30,18){2}{\circle{6}}

\multiput(206,15)(30,18){7}{+}
\multiput(206,15)(-30,18){7}{+}
\put(206,243){+}

\end{picture}
\caption{ \label{fig-trialg} \sf
The directed triangular lattice with orientation given by the
arrows. The sites with fixed states along the pivot line are
marked by filled circles. The open circles mark one particular
position of the boundary line during the traversing of the
lattice.}
\end{figure}
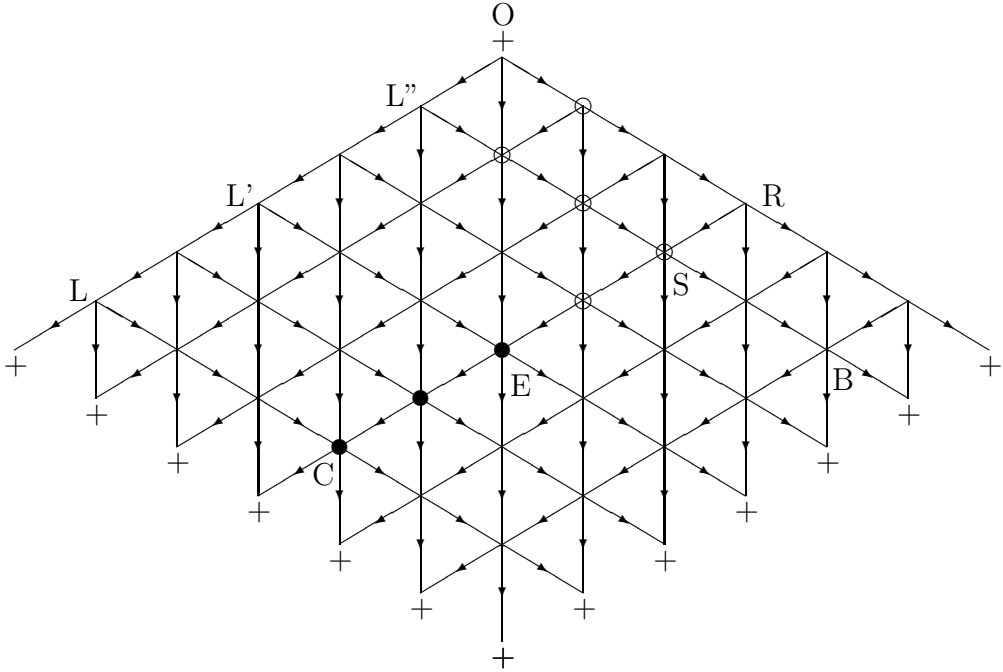

In Figure~\ref{fig-trialg} we show the triangular lattice with
a cut marked by filled circles. In the algorithm the cut is used
as a pivot line by the boundary line which traverse the lattice.
We start by building up the first row at the base CL of the
lattice. We then build up the part of lattice above the cut
from row CL to row EL'. Next the boundary line expands along
the line-piece ES until it reach the position ESL" and the
last site (at L") is flipped to the other site of the top-most
triangle (after this the boundary line is in the position marked
by the open circles). Then we work our way down the right side
of the lattice past R to position ESB. Finally the boundary
line is moved down along the line-piece SEC after which the
whole lattice has been build up. This process is then repeated
for each configuration of the cut. Since the calculations
for different cut-configurations are independent of each other
this algorithm is perfectly suited to take full advantage of
massively parallel computers.

Using this algorithm we calculated $P_N(q)$ for $N \leq 23$ for
the bond and site-bond problems. The integer coefficients of
$P_N(q)$ become very large so the calculation  was performed
using modular arithmetic (see, for example, Knuth 1969).
Each run with $N=23$, using a different moduli, took
approximately 70 hours for the bond problem and 55 hours for
the site-bond using 50 nodes on an Intel Paragon. For the site
problem the weights only depend on whether or not there are
any +'s among the neighbours of the top-most site. As was
the case for the square site problem this may be used to sum
over many configurations of the cut simultaneously (see Jensen
and Guttmann (1995) for further details). This allowed us to
calculate $P_N(q)$ for $N \leq 25$. Each run for $N=25$ took
about 85 hours using 50 nodes.

\section{Extrapolation of the series}

As mentioned, the coefficients of the polynomials
$P_N(q) = \sum_{m \geq 0} a_{N,m} q^m$ will generally agree with
those of the series for $P(q)= \sum_{m \geq 0} a_{m} q^m$ up to
some order, $\tilde{N}$, determined by $N$, but depending on the
specific problem.  In the case of directed bond percolation on
the square lattice Baxter and Guttmann (1988) found that the series
for $P(q)$ could be extended significantly by determining
correction terms to $P_N(q)$. Let us look at

\begin{equation}
P_N - P_{N+1} = q^{\tilde{N}} \sum_{r \geq 0} q^r d_{N,r}
\end{equation}

then we call $d_{N,r} = a_{N,\tilde{N}+r} - a_{N+1,\tilde{N}+r}$
the $r$th correction term.  If formulas can be found for
$d_{N,r}$ for all $r \leq K$ then, using the series coefficients
of $P_N(q)$, one can extend the series for $P(q)$ to order
$\tilde{N}+K$ since

\begin{equation}
a_{\tilde{N}+r} = a_{N,\tilde{N}+r} - \sum_{m=1}^r d_{N+r-m,m}
\end{equation}

for all $r \leq K$. That this method can be very efficient was
demonstrated by Baxter and Guttmann, who identified the first
twelve correction terms for the square bond problem, and used
$P_{29}(q)$ to extend the series for $P(q)$ to order 41. To
really appreciate this advance one should
bear in mind that the time it takes to calculate $P_N(q)$ grows
exponentially with $N$, so a direct calculation correct to the same
order would have taken years rather than days. In the following we
will give details of the correction terms for the various directed
percolation problems on the triangular lattice.

\subsection{The site problem}

For the site problem the coefficients of $P_N(q)$ agree with
those of $P(q)$ to order $N$. In this case the first correction
term is very simple as $d_{N,0} = 2$ for $N \geq 2$, i.e., the
first correction term is simply a constant. For the second
correction term $d_{N,1}$ we find the following sequence

$$
   0,0,3,18,32,50,72,98, \ldots
$$

It is thus immediately clear that

\begin{equation}
  d_{N,1} = 2N^2 \;\; \mbox{ for }\;\; N \geq 3.
\end{equation}

Note that for convenience we assume that the sequence starts
from $N=0$. And indeed we find that for $N \geq r+1$, $d_{N,r}$
can be expressed as a polynomial in $N$ of order $2r$. We have
been able to calculate these polynomials for the the first 10
correction terms. It turns out that it is useful to pull out
a factor $1/(r!(r+1)!)$ and express the correction terms as

\begin{equation}
   d_{N,r} = \frac{1}{r!(r+1)!}
             \sum_{k=0}^{2r} c_r^k N^k. \label{eq:sitecorr}
\end{equation}

This ensures that the coefficients $c_r^k$ in the extrapolation
formulas are integers. We have listed these coefficients in
Table~\ref{table-sitecorr}.

\begin{table}
\scriptsize
\begin{center}
 \begin{tabular}{rrrrrrrrrr} \hline\hline
 \multicolumn{10}{c}{$c_r^k$} \\
\hline
k/r & \multicolumn{1}{c}{1} & \multicolumn{1}{c}{2} &
\multicolumn{1}{c}{3} & \multicolumn{1}{c}{4} &
\multicolumn{1}{c}{5} & \multicolumn{1}{c}{6} &
\multicolumn{1}{c}{7} & \multicolumn{1}{c}{8} &
\multicolumn{1}{c}{9} \\
\hline
0 & 0 & 24 & 0 & 5760 & -345600 & -65318400 &
  -15850598400 & -2984789606400 & -539895767040000\\
1 & 0 & -24 & -48 & -6720 & 662400 & 86728320 &
 15417077760 & 3039204188160 & 681914690150400\\
2 & 4 & 4 & 160 & -2256 & -299136 & -54616320 &
 -10042993152 & -2801552624640 & -758646639912960\\
3 & & -12 & -456 & -5592 & -155040 & 29156640 &
 6930400512 & 1683396497664 & 492391103938560\\
4 & & 8 & 112 & 6968 & 262400 & 3721088 &
 -1895857152 & -641242189440 & -236796916234752\\
5 & & & -72 & -4680 & -211440 & -13781520 &
 -275292864 & 183056948928 & 80349078951936\\
6 & & & 16 & 1016 & 117072 & 9766720 & 775939360
 & 32888441824 & -13942053553664\\
7 & & & & -288 & -35760 & -3900960 & -484442784 &
 -52810790592 & -3002221192320\\
8 & & & & 32 & 6000 & 1183584 & 180360704 &
  27746932192 & 4062978111936\\
9 & & & & & -960 & -222000 & -46002432 &
 -9468263616 & -1860005271168\\
10 & & & & & 64 & 28000 & 8946336 & 2268003232 & 567526218432\\
11 & & & & & & -2880 & -1175328 & -405615168 & -128527251840\\
12 & & & & & & 128 & 112448 & 55739936 & 21947992384\\
13 & & & & & & &  -8064 & -5494272 & -2918143872 \\
14 & & & & & &  & 256 & 406784 & 301743168\\
15 & & & & & & & & -21504 & -23270400\\
16 & & & & & & & & 512 & 1362432\\
17 & & & & & & & & & -55296\\
18 & & & & & & & & & 1024\\
 \hline\hline
 \end{tabular}
 \end{center}
\normalsize
\caption{\label{table-sitecorr} \sf
The coefficients $c_r^k$ in the extrapolation formulas
Eq.~(\protect{\ref{eq:sitecorr}}) for the site problem.}
\end{table}

Obviously since these formulas are correct for $N \geq r+1$
and we have calculated $P_N (q)$ for $N \leq 25$ we did not
have enough terms in the correction sequences to calculate
all the coefficients in these polynomials for the largest
values of $r$. However, from the table of coefficients,
it is immediately clear that $c_r^{2r} = 2^{r+1}$. And in
general we found that $c_r^{2r-m}/2^{r+1}$ is a polynomial
in $r$ of order $2m$

\begin{equation}
   c_r^{2r-m} = \frac{2^{r+1}}{(-4)^m m!}
             \sum_{k=0}^{2m} b_m^k r^k, \label{eq:sitepol}
\end{equation}

where the prefactor has been chosen so as to make the leading
coefficients particularly simple. In Table~\ref{table-sitepol}
we have listed the coefficients $b_m^k$ for the first six
polynomials.

\begin{table}
\small
\begin{center}
 \begin{tabular}{rrrrrrr} \hline\hline
 \multicolumn{7}{c}{$b_m^k$} \\
\hline
k/m & \multicolumn{1}{c}{1} & \multicolumn{1}{c}{2} &
\multicolumn{1}{c}{3} & \multicolumn{1}{c}{4} &
\multicolumn{1}{c}{5} & \multicolumn{1}{c}{6} \\
\hline
1 & 3 & $-3\frac{1}{3}$ & 192 & $4662\frac{2}{5}$
 & -76800 & $2752914\frac{2}{7}$\\
2 & 3 & 19 & -126 & $-20702\frac{2}{3}$ & -969328 & -61888160\\
3 & & $24\frac{2}{3}$ & -411 & 7092 & 1554956
 & $131279844\frac{4}{9}$\\
4 & & 9 & 459 & $21958\frac{1}{3}$ & 196840 & -55417284\\
5 & & & 141 & $-17022\frac{2}{5}$ & -1359655 & $-81930639\frac{1}{3}$\\
6 & & & 27 & $4615\frac{1}{3}$ & 860155 & 105874935\\
7 & & & & 684 & -236446 & $-52835386\frac{20}{21}$\\
8 & & & & 81 & 33050 & 14159255\\
9 & & & & & 3015 & $-2180338\frac{4}{9}$\\
10 & & & & & 243 & 196605\\
11 & & & & & & 12474\\
12 & & & & & & 729\\
 \hline\hline
 \end{tabular}
 \end{center}
\normalsize
\caption{\label{table-sitepol} \sf
The coefficients $b_m^k$ in the extrapolation formulas
Eq.~(\protect{\ref{eq:sitepol}}) for the site problem.}
\end{table}

This time we note that $b_m^{2m} = 3^m$. And indeed as before
we find that $b_m^{2m-j}/3^m$ is a polynomial in $m$ of order
$2j$. In particular we have,

$$
b_m^{2m-1} = 3^m m(17/27+10/27m),
$$
and
$$
b_m^{2m-2} =3^m m(1015/486-5137/1458m+332/243m^2+50/729m^3).
$$

So when calculating the extrapolation formulas Eq.~(\ref{eq:sitecorr})
we first used the sequences for the correction terms to predict
as many polynomials as possible. When we ran out of terms we then
predicted as many of the leading coefficients from
Eq.~(\ref{eq:sitepol}) as possible. This in turn allowed us to find
more extrapolation formulas, which we could use (together with
the formulas for $b_m^{2m-j}$) to find more of the formulas for
$c_k^{2r-m}$. And so on until the process stopped with the 10
extrapolation formulas we listed above.

Using the 10 extrapolation formulas and $P_{25}(q)$ we extended
the series for $P(q)$ through order 35. The resulting series is
listed in Table~\ref{table-siteser}.

\begin{table}
\footnotesize
\begin{center}
 \begin{tabular}{rrrr} \hline\hline
n & $a_n$ & \mbox{\hspace{1cm} n}  & $a_n$ \\
\hline
0   &   1  & 18   &   -111307   \\
1   &   0  & 19   &   -255236   \\
2   &   0  & 20   &   -590543   \\
3   &   -1  & 21   &   -1362919   \\
4   &   -2  & 22   &   -3182137   \\
5   &   -5  & 23   &   -7362611   \\
6   &   -10  & 24   &   -17377129   \\
7   &   -20  & 25   &   -40125851   \\
8   &   -41  & 26   &   -96106251   \\
9   &   -86  & 27   &   -219681825   \\
10   &   -182  & 28   &   -539266908   \\
11   &   -393  & 29   &   -1200140540   \\
12   &   -853  & 30   &   -3087966932   \\
13   &   -1887  & 31   &   -6454135923   \\
14   &   -4208  & 32   &   -18281313306   \\
15   &   -9445  & 33   &   -33072764132   \\
16   &   -21350  & 34   &   -114854030873   \\
17   &   -48612  & 35   &   -145978838818   \\
 \hline\hline
 \end{tabular}
 \end{center}
\normalsize
\caption{\label{table-siteser} \sf
The coefficients $a_n$ in the series expansion of the percolation
probability $P(q)$ for directed site percolation on the triangular
lattice.}
\end{table}

\subsection{The site-bond problem}

For the site-bond problem the coefficients of $P_N (q)$ agree
with those of $P(q)$ to order $N$. In this case the correction
terms are very similar to those of the site problem. In
particular we find that $d_{N,0} = 12$ and in general $d_{N,r}$
is a polynomial in $N$ of order $2r$,

\begin{equation}
   d_{N,r} = \frac{2^r}{r!(r+1)!}
             \sum_{k=0}^{2r} c_r^k N^k. \label{eq:sbcorr}
\end{equation}

We have identified the first 9 correction terms for the
site-bond problem and have listed the coefficients $c_r^k$ in
the extrapolation formulas in Table~\ref{table-sbcorr}.

\begin{table}
\scriptsize
\begin{center}
 \begin{tabular}{rrrrrrrrr} \hline\hline
 \multicolumn{9}{c}{$c_r^k$} \\
\hline
k/r & \multicolumn{1}{c}{1} & \multicolumn{1}{c}{2} &
\multicolumn{1}{c}{3} & \multicolumn{1}{c}{4} &
\multicolumn{1}{c}{5} & \multicolumn{1}{c}{6} &
\multicolumn{1}{c}{7} & \multicolumn{1}{c}{8} \\ \hline
0 & -22 & 372 & -6948 & 228960 & -15136200 &
   1002796200 & -148319942400 & 16196987318400 \\
1 & -28 & -88 & -3570 & 26052 & 532350 & 202151160
   & 54036574200 & 7153213667040 \\
2 & 48 & 66 & 12222 & -66190 & 16300863 &
  -1072631628 & 61870142088 & -28771509693672 \\
3 & & -512 & -6804 & -464344 & -9400240 & -1026322032
   & 27946386678 & 5012953659000 \\
4 & & 192 & 7512 & 428618 & 21649545 & 1760115147 &
   84256658654 & 6746690054058 \\
5 & & & -4800 & -249952 & -23384790 & -1734224880 &
   -194249017018 & -15249026722216 \\
6 & & & 768 & 128960 & 12678024 & 1443885081 &
   172767873502 & 22487197814172 \\
7 & & & & -34816 & -5084160 & -762064416 &
   -111221029556 & -18388293899920 \\
8 & & & & 3072 & 1447680 & 274270176 & 53077387932
  & 10265902430946 \\
9 & & & & & -220160 & -72890880 & -18083074464 &
  -4339851543328 \\
10 & & & & & 12288 & 13020672 & 4539617152 & 1389887209152 \\
11 & & & & & & -1277952 & -833487872 & -335678443520 \\
12 & & & & & & 49152 & 101771264 & 61228145664 \\
13 & & & & & & & -6995968 & -8139063296 \\
14 & & & & & & & 196608 & 721256448 \\
15 & & & & & & & & -36700160 \\
16 & & & & & & & & 786432 \\
 \hline\hline
 \end{tabular}
 \end{center}
\normalsize
\caption{\label{table-sbcorr} \sf
The coefficients $c_r^k$ in the extrapolation formulas
Eq.~(\protect{\ref{eq:sbcorr}}) for the site-bond problem.}
\end{table}

{}From this table it is immediately clear that the coefficient
of the leading order $c_r^{2r} = 3\times4^r$. As in the site
case we find that $c_r^{2r-m}/4^{r+1}$ is a polynomial in $r$ of
order $2m$.

\begin{equation}
   c_r^{2r-m} = \frac{4^{r+1}}{(-4)^m m!}
             \sum_{k=0}^{2m} b_m^k r^k, \label{eq:sbpol}
\end{equation}

where the prefactor has been chosen so as to make the leading
coefficients particularly simple. In Table~\ref{table-sbpol}
we have listed the coefficients $b_m^k$ for the first six
polynomials.

\begin{table}
\small
\begin{center}
 \begin{tabular}{rrrrrr} \hline\hline
 \multicolumn{6}{c}{$b_m^k$} \\
\hline
k/m & \multicolumn{1}{c}{1} &
\multicolumn{1}{c}{2} & \multicolumn{1}{c}{3} &
\multicolumn{1}{c}{4} & \multicolumn{1}{c}{5} \\
\hline
1 & -2 & $-30\frac{1}{2}$ & -177 & $-3187\frac{4}{5}$ & -179760\\
2 & 9 & $3\frac{1}{2}$ & $198\frac{1}{2}$
  & $-3178\frac{1}{2}$ &-101540\\
3 & & -44 & -252 & 3962 & $563989\frac{2}{3}$\\
4 & & 27 & $491\frac{1}{2}$ & $8568\frac{1}{2}$
  & $-153182\frac{1}{2}$\\
5 & & & -342 & $-11196\frac{1}{5}$ & $-381038\frac{1}{3}$\\
6 & & & 81 & 6733 & $401698\frac{1}{2}$\\
7 & & & & -1944 & $-199151\frac{1}{3}$\\
8 & & & & 243 & 57705\\
9 & & & & & -9450\\
10 & & & & & 729\\
 \hline\hline
 \end{tabular}
 \end{center}
\normalsize
\caption{\label{table-sbpol} \sf
The coefficients $b_m^k$ in the extrapolation formulas
Eq.~(\protect{\ref{eq:sbpol}}) for the site-bond problem.}
\end{table}

In this case $b_m^{2m} = 3^{m+1}$ and
$b_m^{2m-1} = 3^{m+1}m(10/27-16/27m)$, which, using the same
procedure as before allowed us to find the first 9 extrapolation
formulas. From $P_{23} (q)$ we were thus able to extend the
series for $P(q)$ through order 32.
The resulting series is listed in Table~\ref{table-sbser}.

\begin{table}
\footnotesize
\begin{center}
 \begin{tabular}{rrrr} \hline\hline
n & $a_n$ & \mbox{\hspace{1cm} n}  & $a_n$ \\
\hline
0   &   1   &   17   &   -86564874   \\
1   &   0   &   18   &   -134834422   \\
2   &   0   &   19   &   -1031059888   \\
3   &   -8   &   20   &   -1842094489   \\
4   &   -4   &   21   &   -12140138712   \\
5   &   -70   &   22   &   -27303542028   \\
6   &   -23   &   23   &   -133912895295   \\
7   &   -640   &   24   &   -447687526744   \\
8   &   -205   &   25   &   -1274069580864   \\
9   &   -6272   &   26   &   -7565668332198   \\
10   &   -2941   &   27   &   -10362711920204   \\
11   &   -64028   &   28   &   -113855530577726   \\
12   &   -47391   &   29   &   -131148651484930   \\
13   &   -678361   &   30   &   -1188175707628214   \\
14   &   -714246   &   31   &   -4485228802915811   \\
15   &   -7495405   &   32   &   1963925987626925   \\
16   &   -10059661   &      &    \\
 \hline\hline
 \end{tabular}
 \end{center}
\normalsize
\caption{\label{table-sbser} \sf
The coefficients $a_n$ in the series expansion of the percolation
probability $P(q)$ for directed site-bond percolation on the triangular
lattice.}
\end{table}

\subsection{The bond problem}

For the bond problem the coefficients of $P_N (q)$ agree with
those of $P(q)$ to order $2N$. In this case the first correction
term is more complicated.  For the first
correction term $d_{N,0}$ we find the following sequence

$$
   1,3,9,27,83,263,857, \ldots
$$

which we have identified as

\begin{equation}
  d_{N,0} = 2C_N-1.
\end{equation}

where $C_N = (2N)!/((N+1)!N!)$ are the Catalan numbers, which also
occur in the correction terms for the square bond problem. In general
we find that for $r \leq 4$ the correction terms are given, for
$N \geq r-2$, by the formulas

\begin{equation}
   d_{N,r} = \sum_{k=1}^{r+1}a_r^k C_{N+k-1} +
             \sum_{k=1}^{r}b_r^k
      \left( \begin{array}{c} N \\ k \end{array} \right)c_{N} +
           \frac{1}{r!r!} \sum_{k=0}^{2r} c_r^k N^k.
      \label{eq:bondcorr}
\end{equation}

We have listed the coefficients $a_r^k$, $b_r^k$ and $c_r^k$ of
these extrapolation formulas in Table~\ref{table-bondcorr}. We note
that as in the previous problems the leading coefficients are
quite simple, $a_r^{r+1} = (-1)^{r}2C_{r+1}$, $b_r^r = 2$, and
$c_r^{2r} = -C_r$.

\begin{table}
\scriptsize
\begin{center}
 \begin{tabular}{rrrrr|rrrr|rrrr} \hline\hline
 \multicolumn{1}{c}{} &
 \multicolumn{4}{c}{$a_r^k$} &
 \multicolumn{4}{c}{$b_r^k$} &
 \multicolumn{4}{c}{$c_r^k$} \\
\hline
k/r &
\multicolumn{1}{c}{1} & \multicolumn{1}{c}{2} &
\multicolumn{1}{c}{3} & \multicolumn{1}{c|}{4} &
\multicolumn{1}{c}{1} & \multicolumn{1}{c}{2} &
\multicolumn{1}{c}{3} & \multicolumn{1}{c|}{4} &
\multicolumn{1}{c}{1} & \multicolumn{1}{c}{2} &
\multicolumn{1}{c}{3} & \multicolumn{1}{c}{4} \\
\hline
0 & & & & & & & & & -1 & -8 & 0 & -2304 \\
1 & 6 & 0 & 52 & -418 & 2 & -12 & 90 & -748 & 2 & 12 & 108 & 1152 \\
2 & -4 & -18 & -56 & 88 & & 2 & -14 & 102 & -1 & -18 & -176 & -1112 \\
3 & & 10 & 72 & 288 & & & 2 & -16 & & 8 & 234 & 2392 \\
4 & & & -28 & -284 & & & & 2 & & -2 & -125 & -3526 \\
5 & & & & 84 & & & & & & & 36 & 2344 \\
6 & & & & & & & & & & & -5 & -820 \\
7 & & & & & & & & & & & & 160 \\
8 & & & & & & & & & & & & -14 \\
 \hline\hline
 \end{tabular}
 \end{center}
\normalsize
\caption{\label{table-bondcorr} \sf
The coefficients $a_r^k$, $b_r^k$ and $c_r^k$ in the extrapolation
formulas Eq.~(\protect{\ref{eq:bondcorr}}) for the bond problem.}
\end{table}

These 5 extrapolation formulas and $P_{23} (q)$ allowed us to
extend the series for $P(q)$ through order 51.
The resulting series is listed in Table~\ref{table-bondser}.

\begin{table}
\footnotesize
\begin{center}
 \begin{tabular}{rrrr} \hline\hline
n & $a_n$ & \mbox{\hspace{1cm} n}  & $a_n$ \\
\hline
0   &   1   &   26   &   1587391   \\
1   &   0   &   27   &   -3535398   \\
2   &   0   &   28   &   6108103   \\
3   &   -1   &   29   &   -13373929   \\
4   &   0   &   30   &   23438144   \\
5   &   -3   &   31   &   -50592067   \\
6   &   1   &   32   &   89703467   \\
7   &   -9   &   33   &   -191306745   \\
8   &   6   &   34   &   342473589   \\
9   &   -29   &   35   &   -722890515   \\
10   &   27   &   36   &   1304446379   \\
11   &   -99   &   37   &   -2729084244   \\
12   &   112   &   38   &   4957423139   \\
13   &   -351   &   39   &   -10292036449   \\
14   &   450   &   40   &   18800279417   \\
15   &   -1275   &   41   &   -38769381587   \\
16   &   1782   &   42   &   71154482443   \\
17   &   -4704   &   43   &   -145869275322   \\
18   &   6998   &   44   &   268798182822   \\
19   &   -17531   &   45   &   -548189750051   \\
20   &   27324   &   46   &   1013680069047   \\
21   &   -65758   &   47   &   -2057857140279   \\
22   &   106211   &   48   &   3816820768061   \\
23   &   -247669   &   49   &   -7717195669953   \\
24   &   411291   &   50   &   14352037073232   \\
25   &   -935107   &   51   &   -28915083150931   \\
 \hline\hline
 \end{tabular}
 \end{center}
\normalsize
\caption{\label{table-bondser} \sf
The coefficients $a_n$ in the series expansion of the percolation
probability $P(q)$ for directed bond percolation on the triangular
lattice.}
\end{table}

\section{Analysis of the series}

We expect that the series for the percolation probability behaves
like

\begin{equation}
P(q) \sim A (1-q/q_c)^{\beta}[1+a_{\Delta}
            (1-q/q_c)^{\Delta} + \ldots ],
\label{eq:crit}
\end{equation}

where $A$ is the critical amplitude, $\Delta$ the leading confluent
exponent and the $\ldots$ represents higher order correction terms.
In the following sections we present the results of our analysis
of the series which include accurate estimates for the critical
parameters $q_c$, $\beta$, $A$ and $\Delta$. For the most part
the best results are obtained using Dlog Pad\'{e} (or in some cases
just ordinary Pad\'{e}) approximants.  A comprehensive review of
these and other techniques for series analysis may be found in
Guttmann (1989).

\subsection{$q_c$ and $\beta$}

In Table~\ref{table-siteana} we list various Dlog Pad\'{e}
approximants to the percolation probability series for directed
site percolation on the triangular lattice.
The defective approximants, those for which there is a spurious
singularity on the positive real axis closer to the origin than
the physical critical point, are marked with an asterisk. Most
higher-order approximants yield estimates around the values
$q_c = 0.4043528$ and $\beta = 0.27645$, with very little spread
among the approximants. Opting for a conservative error estimate,
it seems appropriate to estimate that the critical parameters lie
in the ranges,
$q_c = 0.4043528(10)$ and $\beta = 0.27645(10)$, where the figures
in parenthesis indicate the estimated error on the last digits.

\begin{table}
\small
\begin{center}
 \begin{tabular}{||r|ll|ll|ll||} \hline\hline
 \multicolumn{1}{||r}{ N} &
 \multicolumn{2}{|c}{ [N-1,N]} &
 \multicolumn{2}{|c}{ [N,N]} &
 \multicolumn{2}{|c||}{ [N+1,N]}\\ \hline
 \multicolumn{1}{||r}{} &
 \multicolumn{1}{|c}{$q_c$     } &
 \multicolumn{1}{c}{$\beta$   } &
 \multicolumn{1}{|c}{$q_c$     } &
 \multicolumn{1}{c}{$\beta$   } &
 \multicolumn{1}{|c}{$q_c$     } &
 \multicolumn{1}{c||}{$\beta$   } \\ \hline
  5 & 0.4040928  & 0.27451  & 0.4034610  & 0.27045  & 0.4045236  & 0.27822  \\
  6 & 0.4038500  & 0.27301  & 0.4074251  & 0.31368  & 0.4048775  & 0.28115  \\
  7 & 0.4043787  & 0.27671  & 0.4043331  & 0.27633  & 0.4043677  & 0.27664  \\
  8 & 0.4043535  & 0.27651  & 0.4043803  & 0.27676  & 0.4043698  & 0.27666  \\
  9 & 0.4043615  & 0.27658  & 0.4043636  & 0.27660  & 0.4043555  & 0.27650  \\
 10 & 0.4043623  & 0.27658  & 0.4043582  & 0.27654  & 0.4043574  & 0.27653  \\
 11 & 0.4043567  & 0.27652  & 0.4043567  & 0.27652  & 0.4043576* & 0.27653* \\
 12 & 0.4043567* & 0.27652* & 0.4043610* & 0.27656* & 0.4043553  & 0.27650  \\
 13 & 0.4043525  & 0.27644  & 0.4043538  & 0.27647  & 0.4043580* & 0.27653* \\
 14 & 0.4043529  & 0.27645  & 0.4043526  & 0.27645  & 0.4043528  & 0.27645  \\
 15 & 0.4043527  & 0.27645  & 0.4043529  & 0.27645  & 0.4043528  & 0.27645  \\
 16 & 0.4043528  & 0.27645  & 0.4043528  & 0.27645  & 0.4043528  & 0.27645  \\
 17 & 0.4043528  & 0.27645  &  &  &  &   \\
 \hline\hline
 \end{tabular}
 \end{center}
 \normalsize
\caption{\label{table-siteana} \sf Dlog Pad\'{e} approximants
to the percolation series for directed site percolation on
the triangular lattice.}
\end{table}

The results of the analysis of the series for the bond problem
are listed in Table~\ref{table-bondana}. In this case the spread
among the various approximants is quite substantial, there
appears to be a marked downward drift in the estimates for both
$q_c$ and $\beta$, and the estimates do not settle down to
definite values. It does however seem likely
that the true critical parameters lie within the estimates:
$q_c = 0.52198(1)$ and $\beta = 0.2769(10)$.

\begin{table}
\small
 \begin{center}
 \begin{tabular}{||r|ll|ll|ll||} \hline\hline
 \multicolumn{1}{||r}{ N} &
 \multicolumn{2}{|c}{ [N-1,N]} &
 \multicolumn{2}{|c}{ [N,N]} &
 \multicolumn{2}{|c||}{ [N+1,N]}\\ \hline
 \multicolumn{1}{||r}{} &
 \multicolumn{1}{|c}{$q_c$     } &
 \multicolumn{1}{c}{$\beta$   } &
 \multicolumn{1}{|c}{$q_c$     } &
 \multicolumn{1}{c}{$\beta$   } &
 \multicolumn{1}{|c}{$q_c$     } &
 \multicolumn{1}{c||}{$\beta$   } \\ \hline
 10 & 0.5222235* & 0.28059* & 0.5241918* & 0.25876* & 0.5220853  & 0.27898  \\
 11 & 0.5221835  & 0.28019  & 0.5221078  & 0.27927  & 0.5220958  & 0.27912  \\
 12 & 0.5220691  & 0.27873  & 0.5220388  & 0.27823  & 0.5218366  & 0.27295  \\
 13 & 0.5221336* & 0.27948* & 0.5219680  & 0.27678  & 0.5222844* & 0.28038* \\
 14 & 0.5220278  & 0.27805  & 0.5220029  & 0.27755  & 0.5220086  & 0.27768  \\
 15 & 0.5220076  & 0.27765  & 0.5220064  & 0.27763  & 0.5219973  & 0.27741  \\
 16 & 0.5220101* & 0.27770* & 0.5219613  & 0.27616  & 0.5219942  & 0.27733  \\
 17 & 0.5220046  & 0.27759  & 0.5219895  & 0.27720  & 0.5219959* & 0.27738* \\
 18 & 0.5220774* & 0.27768* & 0.5218335  & 0.26612  & 0.5219770  & 0.27679  \\
 19 & 0.5220382* & 0.27801* & 0.5219944  & 0.27735  & 0.5219876  & 0.27715  \\
 20 & 0.5219795  & 0.27687  & 0.5219848  & 0.27706  & 0.5219846  & 0.27705  \\
 21 & 0.5219846  & 0.27705  & 0.5219848  & 0.27705  & 0.5219847  & 0.27705  \\
 22 & 0.5219847  & 0.27705  & 0.5219848* & 0.27705* & 0.5219780  & 0.27678  \\
 23 & 0.5219837  & 0.27702  & 0.5219820  & 0.27696  & 0.5219811  & 0.27692  \\
 24 & 0.5219767  & 0.27671  & 0.5219804  & 0.27689  & 0.5219830* & 0.27699* \\
 25 & 0.5219796  & 0.27686  & 0.5219827* & 0.27698* & &  \\
 \hline\hline
 \end{tabular}
 \end{center}
 \normalsize
\caption{\label{table-bondana} \sf Dlog Pad\'{e} approximants
to the percolation series for directed bond percolation on
the triangular lattice.}
\end{table}

The analysis of the series for the site-bond problem yields the
results in Table~\ref{table-sbana}. Again we see a downward
drift in the estimates for both $q_c$ and $\beta$ though the
estimates are somewhat more stable than in the previous case.
We estimate that the true critical parameters lie within
the ranges: $q_c = 0.264173(3)$ and $\beta = 0.2766(3)$

\begin{table}
\begin{center}
\small
 \begin{tabular}{||r|ll|ll|ll||} \hline\hline
 \multicolumn{1}{||r}{ N} &
 \multicolumn{2}{|c}{ [N-1,N]} &
 \multicolumn{2}{|c}{ [N,N]} &
 \multicolumn{2}{|c||}{ [N+1,N]}\\ \hline
 \multicolumn{1}{||r}{} &
 \multicolumn{1}{|c}{$q_c$   } &
 \multicolumn{1}{c}{$\beta$  } &
 \multicolumn{1}{|c}{$q_c$   } &
 \multicolumn{1}{c}{$\beta$  } &
 \multicolumn{1}{|c}{$q_c$   } &
 \multicolumn{1}{c||}{$\beta$  } \\ \hline
5 & 0.2639552 & 0.27456 & 0.2639775 & 0.27475 & 0.2645066 & 0.28077\\
6 & 0.2647846 & 0.28559 & 0.2640753 & 0.27556 & 0.2641622 & 0.27647 \\
7 & 0.2641695 & 0.27656 & 0.2641494 & 0.27632 & 0.2641560 & 0.27640 \\
8 & 0.2641576 & 0.27642 & 0.2642476 & 0.27835 & 0.2641667 & 0.27654 \\
9 & 0.2641679 & 0.27655 & 0.2641739 & 0.27665 & 0.2641747 & 0.27666 \\
10 & 0.2641747 & 0.27666 & 0.2641734* & 0.27664* & 0.2641757 & 0.27668 \\
11 & 0.2641758 & 0.27668 & 0.2641753 & 0.27667 & 0.2641754 & 0.27667 \\
12 & 0.2641754 & 0.27667 & 0.2641753 & 0.27667 & 0.2641755* & 0.27668* \\
13 & 0.2641755* & 0.27668* & 0.2641754* & 0.27668* & 0.2641755* & 0.27668*\\þ
14 & 0.2641755* & 0.27668* & 0.2641750 & 0.27667 & 0.2641716 & 0.27654 \\
15 & 0.2641724 & 0.27658 & 0.2641735 & 0.27663 & 0.2641726 & 0.27659 \\
16 & 0.2641729 & 0.27660 & & & & \\
 \hline\hline
 \end{tabular}
 \end{center}
 \normalsize
\caption{\label{table-sbana}\sf Dlog Pad\'{e} approximants
to the percolation series for directed site-bond
percolation on the triangular lattice.}
\end{table}

\subsection{The critical amplitudes}

We can estimate the critical amplitude $A$ by evaluating
Pad\'{e} approximants to $G(q)=(q_c-q)P^{-1/\beta}$ at $q_c$,
since it follows from the leading critical behaviour in
Eq.~(\ref{eq:crit}) that $G(q_c) \sim A^{-1/\beta}q_c$.
This prodecure works well but requires knowledge of both $q_c$
and $\beta$. As we have just shown, we know both $q_c$ and
$\beta$ very accurately for the triangular site series. We estimated
$A$ using values of $q_c$ between 0.4043524 and 0.4043534 and
values of $\beta$ ranging from 0.2764 to 0.2765. For each
$(q_c, \beta)$ pair we calculate $A$ as the average over all
$[N+K,N]$ Pad\'{e} approximants with $K=0,\pm 1$ and
$2N+K \geq 25$. The spread among the approximants is minimal
for $q_c = 0.4043527$, $\beta =0.27645$ where $A = 1.581883(5)$.
Allowing for values of $q_c$ and $\beta$ within the full range
we get $A = 1.5819(4)$.

For the bond problem we used values of $q_c$ from 0.52196 to
0.52121 and $\beta$ from 0.2763 to 0.2773 averaging over Pad\'{e}
approximants with $2N+K \geq 40$. In this case the spread is
minimal for $q_c =0.521985$, $\beta =0.2767$ where
$A = 1.48584(2)$. Again allowing for a wider choice of critical
parameters we estimate that $A = 1.486(6)$.

For the site-bond series we restricted $q_c$ to lie between
0.264170 and 0.264176 and $\beta$ between 0.2763 to 0.2768 using
all approximants with $2N+K \geq 25$. The minimal spread occurs
at $q_c =0.264173$, $\beta =0.2766$ where $A = 1.477393(4)$.
A wider choice for $q_c$ and $\beta$ leads to the estimate
$A = 1.477(1)$.

\subsection{The confluent exponent}

We studied the series using two different methods in order
to estimate the value of the confluent exponent. In the first
method, due to Baker and Hunter (1973), one transforms the
function $P$,
\begin{equation}
P(q) = \sum_{i=1}^{n}A_{i} (1-q/q_c)^{-\lambda_i}
= \sum_{n=0}^{\infty}a_{n}q^{n}
\end{equation}

into an auxiliary function with simple poles at $1/\lambda_{i}$.
We first make the change of variable $q = q_{c}(1-e^{-\zeta})$
and find, after multiplying the coefficient of $\zeta^{k}$ by
$k!$, the auxiliary function

\begin{equation}
{\cal F}(\zeta) = \sum_{i=1}^{N}\sum_{k=0}^{\infty}
A_{i}(\lambda_{i}\zeta)^{k} =
\sum_{i=1}^{N}\frac{A_{i}}{1-\lambda_{i}\zeta},
\end{equation}

which has poles at $\zeta = 1/\lambda_{i}$ with residue
$-A_i/\lambda_i$. The great advantage of
this method is that one obtains simultaneous
estimates for many critical parameters, namely, $\beta$
(the dominant singularity), $\Delta$ (the sub-dominant singularity),
and the critical amplitudes (the residues at the singularities),
while there is only one parameter $q_c$ in the transformation.
Unfortunately this method does not appear to work well for this
problem.
For the site problem we find that the transformed
series generally yields poor estimates for $\beta$ and no estimates
for the confluent exponent. For the bond and site-bond problem
the situation is somewhat better. In Table~\ref{table-bht} we
have listed estimates for the critical parameters
obtained from various Pad\'{e} approximants to the Baker-Hunter
transformed series, using the values $q_c = 0.52198$ for the
bond series and $q_c=0.264173$ for the site-bond series.

\begin{table}[h]
\footnotesize
\begin{center}
 \begin{tabular}{||ccllll||} \hline\hline
 \multicolumn{6}{||c||}{Bond problem} \\ \hline
 \multicolumn{1}{||c}{ N} &
 \multicolumn{1}{c}{ M} &
 \multicolumn{1}{c}{$\beta$} &
 \multicolumn{1}{c}{$A$} &
 \multicolumn{1}{c}{$\Delta$} &
 \multicolumn{1}{c||}{$A\times a_{\Delta}$} \\ \hline
 18 & 19 &    0.27662 &   1.48469 &    1.03897 &   2.21646 \\
 19 & 20 &    0.27705 &   1.48845 &    0.97124 &   1.81301 \\
 20 & 21 &    0.27678 &   1.48604 &    1.01327 &   2.04400 \\
 21 & 21 &    0.28038 &   1.49843 &    0.91120 &   1.68564 \\
 21 & 22 &    0.27677 &   1.48594 &    1.01530 &   2.05671 \\
 22 & 22 &    0.27673 &   1.48582 &    1.01656 &   2.06289 \\
 22 & 23 &    0.27677 &   1.48594 &    1.01530 &   2.05672 \\
 23 & 23 &    0.27559 &   1.48208 &    1.06473 &   2.34714 \\
 23 & 24 &    0.27676 &   1.48587 &    1.01657 &   2.06477 \\
 24 & 25 &    0.27680 &   1.48619 &    1.01064 &   2.02788 \\
 25 & 26 &    0.27679 &   1.48615 &    1.01133 &   2.03211 \\
 \hline
 \multicolumn{6}{||c||}{Site--Bond problem} \\ \hline
 11 & 12 &    0.27788 &   1.48749 &    0.89858 &   1.62193 \\
 12 & 13 &    0.27651 &   1.47668 &    1.01068 &   2.16827 \\
 13 & 13 &    0.27342 &   1.46940 &    1.11395 &   3.15155 \\
 13 & 14 &    0.27651 &   1.47666 &    1.01091 &   2.16997 \\
 14 & 15 &    0.27661 &   1.47745 &    0.99950 &   2.08954 \\
 15 & 15 &    0.27828 &   1.48182 &    0.96056 &   1.91013 \\
 15 & 16 &    0.27659 &   1.47728 &    1.00194 &   2.10641 \\
 \hline\hline
 \end{tabular}
 \end{center}
\normalsize
\caption{\label{table-bht} \sf
The critical exponent $\beta$, confluent exponent $\Delta$ and
critical amplitudes $A$ and $a_{\Delta}$ obtained from $[N,M]$ Pad\'{e}
approximants to the Baker-Hunter transformed series for the bond and
site-bond problems.}
\end{table}

\pagebreak[4]

It should
be noted that, obviously, all approximants yield estimates for
the critical parameters. However, we have discarded many
approximants from the table because we believe the results to
be spurious. For all the discarded approximants we
found that the amplitude of the confluent term was of order zero
and generally the estimate for $\beta$ was very far from the
expected value. Among the remaining approximants we clearly see
that the favoured value of the confluent exponent is $\Delta = 1$.
We also note that the amplitude estimates are in full agreement
with those of the previous section.

In the second method, due to Adler {\em et al.} (1981), one studies
Dlog Pad\'{e} approximants to the function $F(q)$, where

$$
F(q) = \beta P(q) + (q_{c}-q)\mbox{d}P(q)/\mbox{d}q.
$$

The logarithmic derivative of $F(q)$ has a pole at $q_c$ with
residue $\beta + \Delta$. We evaluate the Dlog Pad\'{e} approximants
for a range of values of $q_c$ and $\beta$. In
Table~\ref{table-ampt} we have listed the estimates for $\Delta$
obtained by averaging over all $[N,N+K]$ approximants for a few
values of $\beta$ with $q_c$ fixed at the central value of our
estimate range. For the site and site-bond problem we used all
approximants with $2N+K \geq 25$ and for the bond problem all
approximants with $2N+K \geq 40$. This analysis clearly indicates
that $\Delta \simeq 1$ and thus that there is no sign of any
non-analytic corrections to scaling.

\begin{table}[h]
\small
\begin{center}
 \begin{tabular}{||cc|cc|cc||} \hline\hline
  \multicolumn{2}{||c|}{Site problem} &
  \multicolumn{2}{c|}{Site-bond problem} &
  \multicolumn{2}{c||}{Bond problem} \\
  \hline
 $\beta$ & $\Delta$ & $\beta$ & $\Delta$ &$\beta$ & $\Delta$ \\
\hline
0.27640 & 0.98587 & 0.27630 & 0.97076 & 0.27660 & 1.03471 \\
0.27641 & 0.99003 & 0.27635 & 0.98220 & 0.27665 & 1.03079 \\
0.27642 & 0.99378 & 0.27640 & 0.99136 & 0.27670 & 1.02537 \\
0.27643 & 0.99683 & 0.27645 & 0.99796 & 0.27675 & 1.01846 \\
0.27644 & 0.99890 & 0.27650 & 1.00176 & 0.27680 & 1.01013 \\
0.27645 & 0.99979 & 0.27655 & 1.00262 & 0.27685 & 1.00042 \\
0.27646 & 0.99942 & 0.27660 & 1.00047 & 0.27690 & 0.98941 \\
0.27647 & 0.99782 & 0.27665 & 0.99533 & 0.27695 & 0.97716 \\
0.27648 & 0.99514 & 0.27670 & 0.98732 & 0.27700 & 0.96377 \\
0.27649 & 0.99164 & 0.27675 & 0.97663 & 0.27705 & 0.94934 \\
0.27650 & 0.98755 & 0.27680 & 0.96352 & 0.27710 & 0.93394 \\
 \hline\hline
 \end{tabular}
 \end{center}
\normalsize
\caption{\label{table-ampt} \sf
Estimates for the confluent exponent $\Delta$ from the transformation
due to Adler {\em et al.} (1981) for various values of $\beta$
at the critical point $q_c$.}
\end{table}

\pagebreak[4]

\section{Conclusion}

\begin{table}
\small
 \begin{center}
 \begin{tabular}{||l|lll|llc||} \hline\hline
 \multicolumn{1}{||c}{Problem} &
 \multicolumn{3}{|c|}{Unbiased estimates} &
 \multicolumn{3}{c||}{Biased estimates}\\ \hline
 \multicolumn{1}{||r}{} &
 \multicolumn{1}{|c}{$q_c$} &
 \multicolumn{1}{c}{$\beta$} &
 \multicolumn{1}{c}{$A$} &
 \multicolumn{1}{|c}{$q_c$} &
 \multicolumn{1}{c}{$A$} &
 \multicolumn{1}{c||}{$N_{min}$} \\ \hline
T bond & 0.52198(1) & 0.2769(10) & 1.486(6)
& 0.521971(5) &  1.4841(2) & 45 \\
T site & 0.4043528(10) & 0.27645(10) & 1.5819(5)
& 0.4043523(3) & 1.58183(2) & 30 \\
T site-bond & 0.264173(3) & 0.2766(3) & 1.477(1)
& 0.264170(4) & 1.4765(3) & 25 \\
S bond & 0.3552994(10) & 0.27643(10) & 1.3292(5)
& 0.35529955(15) & 1.32920(1) & 45 \\
S site & 0.294515(5) & 0.2763(3) & 1.425(1)
& 0.294518(3) & 1.42588(4) & 30 \\
H bond & 0.177143(2) & 0.2763(2) & 1.106(1)
& 0.177144(2) & 1.1064(3) & 30 \\
H site & 0.160067(5) & 0.2763(4) & 1.167(1)
& 0.160069(2) & 1.1680(3) & 30 \\
 \hline\hline
 \end{tabular}
 \end{center}
 \normalsize
\caption{\label{table-summary} \sf Estimates of
critical parameters for the three problems on the
triangular (T) lattice studied in this paper and
for the site and bond problems
on the directed square (S) and honeycomb (H) lattices.
See the text for explanation of the biased estimates.}
\end{table}

In this paper we have presented extended series for the percolation
probability for site, bond and site-bond percolation on the
directed triangular lattice. The analysis of the series leads to
improved estimates for the percolation threshold and the order
parameter exponent $\beta$. In Table~\ref{table-summary} we
summarise the critical parameter estimates for the percolation
probability for the three problems on the triangular lattice
studied here and the problems studied in our earlier paper.
The estimates for $q_c = 1- p_c$ for the triangular bond and
site problems are in excellent agreement with those obtained
by Essam {\em et al.} (1986, 1988), $q_c =  0.40437(7)$
and $q_c= 0.521975(7)$, respectively. The estimates for
$\beta$ clearly show, as one would expect, that all the models
studied in this and our earlier paper belong to the same
universality class. The unbiased estimates for $\beta$,
derived in the manner described in the previous section,
for the triangular site and square bond cases are in excellent
agreement and have small error bars (we emphasize once more
that our error estimates are conservative). This leads us to
belive that an improved estimate $\beta = 0.27644(3)$ is
reasonable. We used this highly accurate estimate to obtain
the {\em biased} estimates in Table~\ref{table-summary}
as follows. First we formed the series for
$P(q)^{-1/\beta}$ using $\beta = 0.27644$. This series has a
simple pole at $q_c$ which can be estimated from ordinary
Pad\'{e} approximants. By averaging over all $[N,N+K]$
approximants with $K= 0,\pm 1$ and $2N+K \geq N_{min}$
we obtained the biased estimates for $q_c$ the error-bars
are basicly twice the spread among the approximants. We then
used the biased estimate for $q_c$ (with $\beta$ as before)
to obtain the biased estimates for the amplitudes using the
procedure decribed in the previous section.
As previously noted (Jensen and Guttmann 1995), there is no
simple rational
fraction whose decimal expansion agrees with
our estimate of $\beta$. Given
that this model is not conformally
invariant, and that the expectation of exponent
rationality is a consequence of conformal invariance,
it is perhaps naive
to expect otherwise. It is nevertheless true
that there is a widely held
- if imprecisely expressed - view that
two dimensional systems should have
rational exponents. More precise numerical
work such as the recent estimation
of the longitudinal size exponent $\nu_{||}$
(Conway and Guttmann 1994) of
directed animals and
the present calculation, supports the conclusion that the critical
exponents for these models should not be expected
to be simple rational fractions.
Finally note that none of the series show any evidence of
non-analytic confluent correction terms. This provides a hint that
the models might be exactly solvable.

\section*{Acknowledgements}

Financial support from the Australian Research Council is
gratefully acknowledged.

\section*{References}

\refjl{Adler J, Moshe M and Privman V 1981}{\JPA}{17}{2233}

\refjl{Baker G A and Hunter D L 1973}{\PRB}{7}{3377}

\refjl{Baxter R J and Guttmann A J 1988}{\JPA}{21}{3193}

\refjl{Bidaux R and Forgacs G 1984}{\JPA}{17}{1853}

\refjl{Blease J 1977}{\JPC}{10}{917}

\refjl{Bousquet-M\'{e}lou M 1995}{Percolation models and animals}
{}{LaBRI-University of Bourdeaux preprint 1082-1995}

\refjl{Conway A R and Guttmann A G 1994}{\JPA}{27}{7007}

\refjl{De'Bell K and Essam J W 1983}{\JPA}{16}{3145}

\refjl{Essam J W, De'Bell K, Adler J and Bhatti F M 1986}{\PRB}{33}{1982}

\refjl{Essam J W, Guttmann A J and De'Bell K 1988}{\JPA}{21}{3815}

\refbk{Guttmann  A J 1989 Asymptotic Analysis of Power-Series
Expansions}{Phase Transitions and Critical Phenomena}
{vol 13, eds. C. Domb and J. L. Lebowitz (Academic Press, New York).}

\refjl{Jensen I and Guttmann A J 1995}{\JPA}{28}{4813}

\refbk{Knuth D E 1969}{Seminumerical Algorithms (The Art of
Computer Programming 2)}{(Addison-Wesley, Reading MA)}

\newpage

\appendix
\setcounter{equation}{0}
\renewcommand{\theequation}{A.\arabic{equation}}
{\Large \bf Appendix: The first extrapolation formulas}

\vspace{3mm}

In this appendix we shall calculate the first correction
term(s) $d_{N,r}$ for the various problems we have studied
in this paper.  In the following we rely heavily on the work
of Bousquet-M\'{e}lou (1995) and we shall represent the
directed percolation models in terms of directed animals.
By a directed site (bond) animal $A$ we simply understand any
finite set of connected sites (bonds) starting at the
origin O in Figure~\ref{fig-oriented}. The {\em area} (or size)
$|A|$ of an animal is the number of sites in the animal
and the {\em perimeter} $p(A)$ is the number of unoccupied
sites (bonds) with a nearest neighbour in $A$. The {\em height}
$h$ of an animal is the last row to which the animal extends, i.e.,
there is at least one occupied site in row $h$ belonging to $A$
but none in row $h+1$. The percolation probability, for the site
and site-bond cases, is

\begin{equation}
  P(q) = 1 - \sum_{A \in {\cal A}} q^{p(A)}(1-q)^{|A|-1}
\end{equation}

where ${\cal A}$ denotes the set of animals on the lattice.
For bond percolation the power of $(1-q)$ in the above equation
is $|A|$. The difference stems from the assumption that for site
percolation the origin is occupied with probability 1.
In analogy with the finite-lattice formulation we define
subsets ${\cal A}_N$ of ${\cal A}$ as the set of animals
of height less than $N$. It follows that

\begin{equation}
  P_N(q) = 1 - \sum_{A \in {\cal A}_N} q^{p(A)}(1-q)^{|A|-1},
\end{equation}

and

\begin{equation}
  P_N(q)-P_{N+1}(q)=
           1 - \sum_{A \in {\cal A}_{N+1}\backslash{\cal A}_N}
                     q^{p(A)}(1-q)^{|A|-1}. \label{eq:pndiff}
\end{equation}

It should be noted that in the site and site-bond cases
the polynomials $P_N(q)$ defined above are identical to the
polynomials $P_{N+1}(q)$ from Section~2. From Eq.~(\ref{eq:pndiff})
it is immediately clear that $P_N$ and $P_{N+1}$ agree up to
an order $\tilde{N}$ determined by the animals in
${\cal A}_{N+1}\backslash {\cal A}_N$ with the smallest perimeter.
In our cases $\tilde{N}$ is simply proportional to $N$ and the
polynomials $P_N(q)$ therefore have a formal limit $P_{\infty}(q)$
which we identify as the percolation probability $P(q)$.
By expanding Eq.~(\ref{eq:pndiff}) one gets a very useful
expression for the correction terms

\begin{equation}
  P_N(q)-P_{N+1}(q)= q^{\tilde{N}}\sum_{r \geq 0} q^r d_{N,r} =
  q^{\tilde{N}}\sum_{r \geq 0} q^r
  \sum_{k=0}^{r} \sum_{{\cal A} \in {\cal A}_{N,k}}
      (-1)^{r-k}\left( \begin{array}{c} |A|-1 \\ r-k \end{array}
       \right), \label{eq:corrani}
\end{equation}

where ${\cal A}_{N,k} = \{ A \in {\cal A}_{N+1}\backslash
{\cal A}_N,\;\; p(A) = \tilde{N}+k \}$.

{\large \bf The site case}

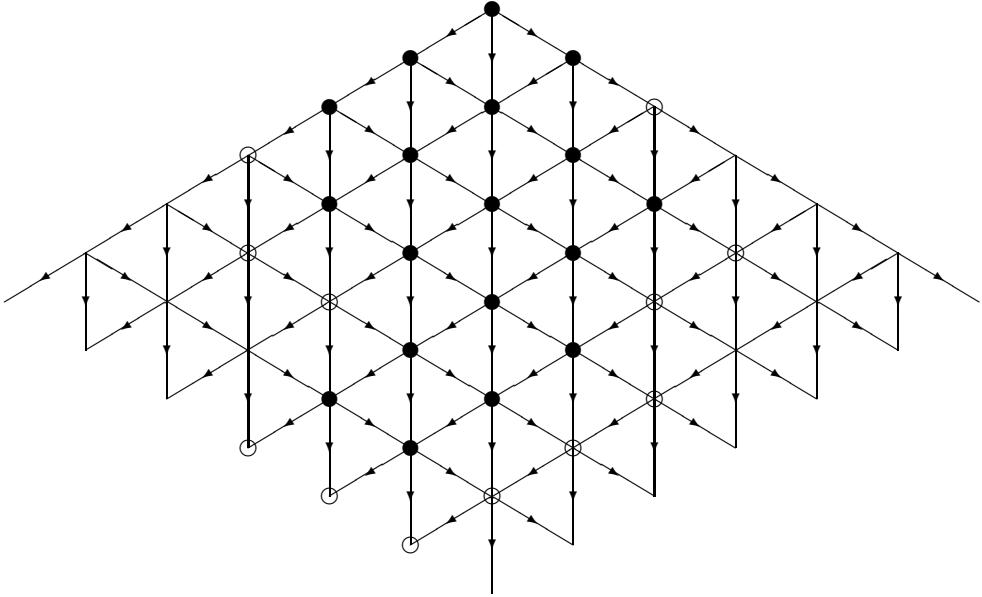
\begin{figure}[tb]
\begin{picture}(420,260)

\multiput(240,42)(30,18){6}{\line(-5,3){180}}
\multiput(180,42)(-30,18){6}{\line(5,3){180}}
\multiput(210,24)(-30,18){6}
{\multiput(0,0)(30,18){6}{\line(0,1){36}}}

\multiput(60,132)(30,-18){6}
{\multiput(0,0)(30,18){6}{\vector(0,-1){2}}}
\multiput(45,141)(30,-18){6}
{\multiput(0,0)(30,18){6}{\vector(-3,-2){2}}}
\multiput(75,141)(30,-18){6}
{\multiput(0,0)(30,18){6}{\vector(3,-2){2}}}

\multiput(210,240)(30,-18){2}{\circle*{6}}
\multiput(180,222)(30,-18){4}{\circle*{6}}
\multiput(150,204)(30,-18){4}{\circle*{6}}
\multiput(150,168)(30,-18){4}{\circle*{6}}
\multiput(180,114)(30,-18){2}{\circle*{6}}
\multiput(150,96)(30,-18){2}{\circle*{6}}

\multiput(270,204)(-30,-18){1}{\circle{6}}
\multiput(300,150)(-30,-18){2}{\circle{6}}
\multiput(270,96)(-30,-18){3}{\circle{6}}
\multiput(120,186)(30,-18){1}{\circle{6}}
\multiput(120,150)(30,-18){2}{\circle{6}}
\multiput(120,78)(30,-18){3}{\circle{6}}

\end{picture}
\caption{ \label{fig-triani} \sf
A compact directed site animal (filled circles) on the triangular
lattice with perimeter sites marked by open circles.}
\end{figure}

An animal is {\em compact} if the occupied sites in any given row
are consecutive,  i.e., there are no holes in the animal (see
Figure~\ref{fig-triani}). Obviously, removing interior sites from
a compact animal can never reduce the perimeter. Therefore, the
animals in ${\cal A}_{N+1}\backslash {\cal A}_N$ with minimal
perimeter are compact. The minimal perimeter of a compact animal
of height $N$ is $N+2$. This is proved by induction on $N$.
It is obviously true for $N=1$ and one can easily see that by
adding sites in row $N+1$ to a compact animal of height $N$
at least one more perimeter-site is added. We also note that there
are at least two animals of height $N$ with perimeter $N+2$, namely
a string of sites (one per row) running down either the left or
right hand side of the lattice. This shows that $\tilde{N} = N+2$.
It is also clear that these two animals must be the ones that give
rise to the first correction term $d_{N,0} = 2$. What remains is
to prove that there can be no more animals in
${\cal A}_{N+1}\backslash {\cal A}_N$ with perimeter $N+2$.
In order to do this we need a unique way of characterising the
perimeter of compact animals of height $N$. Introduce lines
$R_k$ ($L_k$) parallel to the right-hand (left-hand)
egde starting from row $k$. Since the animal is compact all sites
in $A$ intersecting $R_k$ and $L_k$ are consecutive. The number
of perimeter sites on the left-hand side of the animal is
$w_l = \max \{k, L_k \cap A \neq \emptyset \}$ because the
last occupied site in line $L_k$ has an unoccupied
neighbour on $L_k$. Similar arguments apply for the number
of perimeter sites $w_r$ on the right side of $A$. Finally
we note that the only perimeter site not accounted for is
the one lying vertically below the last site in $L_N$ and/or
$R_N$. So the perimeter is $p(A) = w_l+w_r+1$. Furthermore
if $A \in {\cal A}_{N+1}\backslash {\cal A}_N$ then either
$w_l$ or $w_r$ (possibly both) has to equal $N$. The animals
with minimal perimeter $N+2$ are those with $w_l=1$ or $w_r=1$,
obviously there can be only two such animals, which completes
our proof that $d_{N,0} = 2$.

{}From Eq.~(\ref{eq:corrani}) we get the second correction term

\begin{equation}
d_{N,1} = |{\cal A}_{N,1}| - \sum_{A \in {\cal A}_{N,0}}(|A|-1),
 \label{eq:tscorr2}
\end{equation}

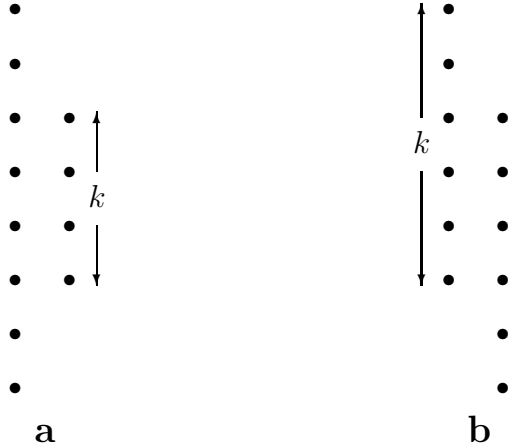
\begin{figure}[tb]
\begin{picture}(400,180)

\multiput(140,20)(0,20){8}{\circle*{4}}
\multiput(160,60)(0,20){4}{\circle*{4}}
\put(147,0){\large \bf a}
\put(167,87){$k$}
\put(170,100){\vector(0,1){22}}
\put(170,80){\vector(0,-1){22}}

\multiput(300,60)(0,20){6}{\circle*{4}}
\multiput(320,20)(0,20){6}{\circle*{4}}
\put(307,0){\large \bf b}
\put(287,107){$k$}
\put(290,120){\vector(0,1){42}}
\put(290,100){\vector(0,-1){42}}

\end{picture}
\caption{ \label{fig-tricor2} \sf
The two types of compact directed site animals with $w_l = 2$ which
contribute to the second correction term.}
\end{figure}

where $|{\cal A}_{N,1}|$ is the number of animals of height $N$
with a perimeter of length $N+3$. From the characterisation of
compact animals derived above it follows that the animals in
${\cal A}_{N,1}$ are those with $w_l=2$ or $w_r=2$. Obviously
there is the same number of animals in each case so we can
restrict ourselves to the case $w_l=2$, $w_r=N$. We are thus
looking at animals restricted to the left-most two lines
$L_1$ and $L_2$ of the lattice and either $L_1 \cap R_N$ or
$L_2 \cap R_N$ has to be non-empty. The two types of animals are
illustrated in Figure~\ref{fig-tricor2}. If
$L_1 \cap R_N \neq \emptyset$ (Figure~\ref{fig-tricor2}a)
then the first $N$ sites
of $L_1$ are occupied and $1 \leq k \leq N$ consecutive sites
along $L_2$ are occupied; these $k$ sites can be placed in
$N-k+1$ positions. If $L_1 \cap R_N = \emptyset$
(Figure~\ref{fig-tricor2}b) and the
first $k$ sites ($1 \leq k \leq N-1$) of $L_1$ are occupied then
the first $j$ consecutive sites $0 \leq j \leq k$ of $L_2$ may
be empty. Combining these two contributions with those from
$w_r =2$ we find

$$
|{\cal A}_{N,1}| = 2\left( \sum_{k=1}^N (N-k+1) +
 \sum_{k=1}^{N-1} (k+1) \right) = 2N^2 + 2N - 2.
$$

Since the number of sites in each of the two animals in
${\cal A}_{N,0}$ is $N$, Eq.~(\ref{eq:tscorr2}) yields

$$
d_{N,1} = 2N^2
$$

thus proving the empirical formula derived previously.

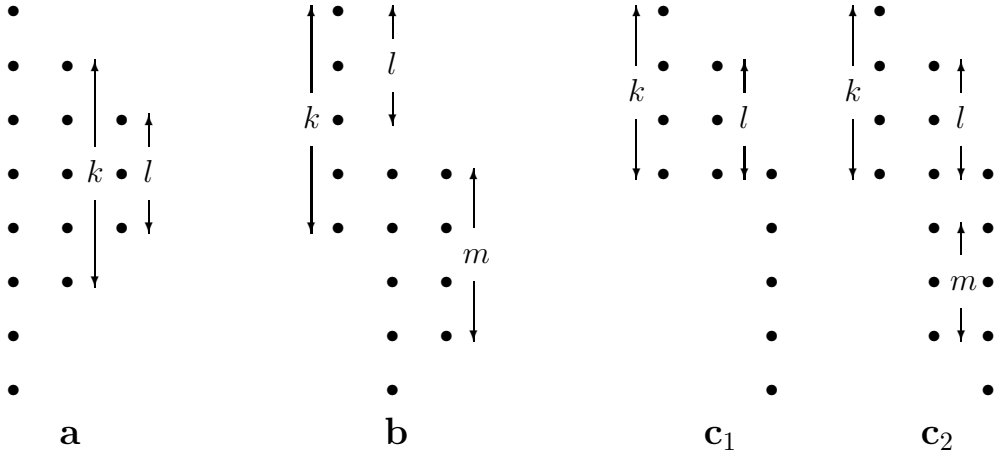
\begin{figure}[tb]
\begin{picture}(400,180)

\multiput(20,20)(0,20){8}{\circle*{4}}
\multiput(40,60)(0,20){5}{\circle*{4}}
\multiput(60,80)(0,20){3}{\circle*{4}}
\put(37,0){\large \bf a}
\put(47,97){$k$}
\put(50,110){\vector(0,1){32}}
\put(50,90){\vector(0,-1){32}}
\put(68,97){$l$}
\put(70,110){\vector(0,1){12}}
\put(70,90){\vector(0,-1){12}}

\multiput(140,80)(0,20){5}{\circle*{4}}
\multiput(160,20)(0,20){5}{\circle*{4}}
\multiput(180,40)(0,20){4}{\circle*{4}}
\put(157,0){\large \bf b}
\put(127,117){$k$}
\put(130,130){\vector(0,1){32}}
\put(130,110){\vector(0,-1){32}}
\put(158,137){$l$}
\put(160,150){\vector(0,1){12}}
\put(160,130){\vector(0,-1){12}}
\put(186,67){$m$}
\put(190,80){\vector(0,1){22}}
\put(190,60){\vector(0,-1){22}}

\multiput(260,100)(0,20){4}{\circle*{4}}
\multiput(280,100)(0,20){3}{\circle*{4}}
\multiput(300,20)(0,20){5}{\circle*{4}}
\put(275,0){\large \bf c$_1$}
\put(247,127){$k$}
\put(250,140){\vector(0,1){22}}
\put(250,120){\vector(0,-1){22}}
\put(288,117){$l$}
\put(290,130){\vector(0,1){12}}
\put(290,110){\vector(0,-1){12}}

\multiput(340,100)(0,20){4}{\circle*{4}}
\multiput(360,40)(0,20){6}{\circle*{4}}
\multiput(380,20)(0,20){5}{\circle*{4}}
\put(355,0){\large \bf c$_2$}
\put(327,127){$k$}
\put(330,140){\vector(0,1){22}}
\put(330,120){\vector(0,-1){22}}
\put(368,117){$l$}
\put(370,130){\vector(0,1){12}}
\put(370,110){\vector(0,-1){12}}
\put(366,57){$m$}
\put(370,70){\vector(0,1){12}}
\put(370,50){\vector(0,-1){12}}

\end{picture}
\caption{ \label{fig-tricor3} \sf
The types of compact directed site animals with $w_l = 3$ which
contributes to the third correction term.}
\end{figure}

Next we prove the formula for $d_{N,2}$. From Eq.~(\ref{eq:corrani})
we see that the third correction term is given by

\begin{equation}
d_{N,2} = |{\cal A}_{N,2}| - \sum_{A \in {\cal A}_{N,1}}(|A|-1)
   + \sum_{A \in {\cal A}_{N,0}}
     \left( \begin{array}{c} |A|-1 \\ 2 \end{array} \right).
 \label{eq:tscorr3}
\end{equation}

In this case there are two distinctly different sets of animals
in ${\cal A}_{N,2}$, namely, compact animals with $w_l = 3$ as
pictured in Figure~\ref{fig-tricor3}, and animals formed from
the compact animals of Figure~\ref{fig-tricor2} by removing
consecutive sites from the second line of occupied sites leaving
at least the first and last sites untouched. One easily sees that
cutting such a 'hole' in these animals is the only way of increasing
their perimeter by one site. From the animals in
Figure~\ref{fig-tricor3} we get the following contributions

\begin{eqnarray}
\mbox{\large \bf a}: &
{\displaystyle 2\sum_{k=1}^N\sum_{l=1}^k(N-k+1)(k-l+1)} = &
\frac{1}{12}N^4+\frac12 N^3 + \frac{11}{12}N^2 + \frac12 N,\nonumber\\
\mbox{\large \bf b}: &
{\displaystyle 2\sum_{k=1}^{N-1}\sum_{l=1}^k\sum_{m=1}^{N-l}(N-l-m+1)}
 = & \frac14 N^4 + \frac56 N^3 - \frac14 N^2 - \frac56 N,\nonumber \\
\mbox{\large \bf c$_1$}: &
{\displaystyle 2\sum_{k=1}^{N-1}\sum_{l=1}^k (l+1)} = &
\frac13 N^3 + N^2  - \frac43 N,\label{eq:corr31}\\
\mbox{\large \bf c$_2$}: &
{\displaystyle 2\sum_{k=1}^{N-2}\sum_{l=0}^k\sum_{m=1}^{N-k-1}(m+l+1)}
 = & \frac16 N^4 + \frac13 N^3 - \frac76 N^2 - \frac43 N + 2.\nonumber
\end{eqnarray}

The animals in Figure~\ref{fig-tricor3}a account for animals
with $L_1 \cap R_N \neq \emptyset$, those of \ref{fig-tricor3}b
for animals with $L_1 \cap R_N = \emptyset$ and
$L_2 \cap R_N \neq \emptyset$, and lastly those of
\ref{fig-tricor3}c for animals where
$L_1 \cap R_N = \emptyset$ and $L_2 \cap R_N = \emptyset$. The
contribution in each case is simply all the possible configurations
which leads to an animal of the specified kind. The sums in
Eq.~(\ref{eq:corr31}) should be self-evident.

The animals in Figure~\ref{fig-tricor2} with a cut as described above
yield the contributions

\begin{eqnarray}
\mbox{\large \bf a}: &
{\displaystyle 2\sum_{k=3}^N\sum_{l=1}^{k-2}(N-k+1)(k-l-1)} = &
\frac{1}{12}N^4 - \frac16 N^3 - \frac{1}{12}N^2 + \frac16 N, \nonumber\\
\mbox{\large \bf b}: &
{\displaystyle 2\sum_{k=2}^{N-1}\sum_{l=0}^{k-2}
\sum_{m=1}^{k-l-1}(k-l-m)}
 = & \frac{1}{12}N^4 - \frac16 N^3 - \frac{1}{12}N^2 + \frac16 N.
\label{eq:corr32}
\end{eqnarray}

In case (a) the piece in the second line has to have at least three
sites ($k \geq 3$) otherwise one could not cut out a hole of size
$l \leq k-2$. The $k$ sites can be placed in $(N-k+1)$ positions
and the hole can be cut in $k-2-l+1 = k-l-1$ places, which leads
to the first sum. In case (b) there can be from 2 to $N-1$
sites in the first line (the sum over $k$) with an overlap
of $0 \leq m \leq k-2$ sites between the first line and the
consecutive sites in the second line extending to the $N$th row.
Among the remaining $k-m$ sites in the second line
$1 \leq l \leq k-m-1$ are occupied and they can be placed
in $k-m-l$ positions, thus giving us the second sum.

The second term in Eq.~(\ref{eq:tscorr3}) is the sum over
$|A|-1$ of the compact animals in Figure~\ref{fig-tricor2} and
we find the two contributions:

\begin{eqnarray}
\mbox{\large \bf a}: &
{\displaystyle 2\sum_{k=1}^N(N-k+1)(N+k-1)} = &
\frac43 N^3 + N^2 - \frac13 N, \nonumber \\
\mbox{\large \bf b}: &
{\displaystyle 2\sum_{k=1}^{N-1}\sum_{l=0}^{k}(N+k-l-1)} = &
\frac43 N^3 - \frac{10}{3} N + 2.
\label{eq:corr33}
\end{eqnarray}

Finally the last term in Eq.~(\ref{eq:tscorr3}) simply stems from
the two animals in ${\cal A}_{N,0}$ and their contribution is

\begin{eqnarray}
2\left(\begin{array}{c} N-1\\ 2\end{array}\right)& = & N^2-3N+2.
\label{eq:corr34}
\end{eqnarray}

By adding the contributions of Eqs.~(\ref{eq:corr31}),
(\ref{eq:corr32}) and (\ref{eq:corr34}) while subtracting those
of Eq.~(\ref{eq:corr33}) we get

\begin{equation}
  d_{N,2} = \frac23 N^4 - N^3 + \frac13 N^2 - 2N + 2 =
\frac{1}{12}(8N^4 - 12N^3 + 4 N^2 - 24N + 24)
\end{equation}

in full agreement with the extrapolation formula listed in
Table~\ref{table-sitecorr}, thus concluding the proof for
$d_{N,2}$.

\vspace{3mm}

{\large \bf The site-bond case}

\begin{figure}[tb]
\begin{picture}(420,260)
\newsavebox{\swb}
\newsavebox{\seb}
\newsavebox{\ssb}
\newsavebox{\swp}
\newsavebox{\sep}
\newsavebox{\ssp}

\savebox{\swb}(30,18){\thicklines \put(1.6,2.0){\line(5,3){27}}
\put(3,1.2){\line(5,3){27}}}
\savebox{\seb}(30,18){\thicklines \put(-3,1.2){\line(-5,3){27}}
\put(-1.6,2.0){\line(-5,3){27}}}
\savebox{\ssb}(4,36){\thicklines \put(-0.7,2){\line(0,1){32}}
\put(0.7,2){\line(0,1){32}}}

\multiput(240,42)(30,18){6}{\line(-5,3){180}}
\multiput(180,42)(-30,18){6}{\line(5,3){180}}
\multiput(210,24)(-30,18){6}
{\multiput(0,0)(30,18){6}{\line(0,1){36}}}

\multiput(60,132)(30,-18){6}
{\multiput(0,0)(30,18){6}{\vector(0,-1){2}}}
\multiput(45,141)(30,-18){6}
{\multiput(0,0)(30,18){6}{\vector(-3,-2){2}}}
\multiput(75,141)(30,-18){6}
{\multiput(0,0)(30,18){6}{\vector(3,-2){2}}}

\multiput(210,240)(30,-18){2}{\circle*{6}}
\multiput(180,222)(30,-18){4}{\circle*{6}}
\multiput(150,204)(30,-18){4}{\circle*{6}}
\multiput(150,168)(30,-18){4}{\circle*{6}}
\multiput(180,114)(30,-18){2}{\circle*{6}}
\multiput(150,96)(30,-18){2}{\circle*{6}}

\multiput(270,204)(-30,-18){1}{\circle{6}}
\multiput(300,150)(-30,-18){2}{\circle{6}}
\multiput(270,96)(-30,-18){3}{\circle{6}}
\multiput(120,186)(30,-18){1}{\circle{6}}
\multiput(120,150)(30,-18){2}{\circle{6}}
\multiput(120,78)(30,-18){3}{\circle{6}}

\multiput(150,204)(30,18){2}{\usebox{\swb}}
\multiput(150,168)(30,18){3}{\usebox{\swb}}
\multiput(150,132)(30,18){3}{\usebox{\swb}}
\multiput(150,96)(30,18){4}{\usebox{\swb}}
\multiput(150,60)(30,18){3}{\usebox{\swb}}
\put(240,222){\usebox{\seb}}
\multiput(270,168)(-30,18){3}{\usebox{\seb}}
\multiput(270,132)(-30,18){4}{\usebox{\seb}}
\multiput(240,114)(-30,18){3}{\usebox{\seb}}
\multiput(240,78)(-30,18){2}{\usebox{\seb}}
\multiput(210,60)(-30,18){2}{\usebox{\seb}}
\put(150,60){\usebox{\ssb}}
\multiput(150,132)(0,36){2}{\usebox{\ssb}}
\multiput(180,78)(0,36){4}{\usebox{\ssb}}
\multiput(210,60)(0,36){5}{\usebox{\ssb}}
\multiput(240,78)(0,36){4}{\usebox{\ssb}}
\put(270,132){\usebox{\ssb}}

\put(128,191){$\parallel$}
\put(128,155){$\parallel$}
\put(128,83){$\parallel$}
\put(176,50){=}
\put(256,101){$\parallel$}
\put(286,155){$\parallel$}
\put(256,209){$\parallel$}

\end{picture}
\caption{ \label{fig-tsbani} \sf
A site-compact directed site-bond animal (filled circles and thick
bonds) on the triangular
lattice with possible perimeter sites marked by open circles.
Some of the perimeter sites have only one possible incident
bond (marked by double lines) and in those cases the bond can
be present (the site is part of the perimeter) or absent (the edge
is part of the perimeter).}
\end{figure}
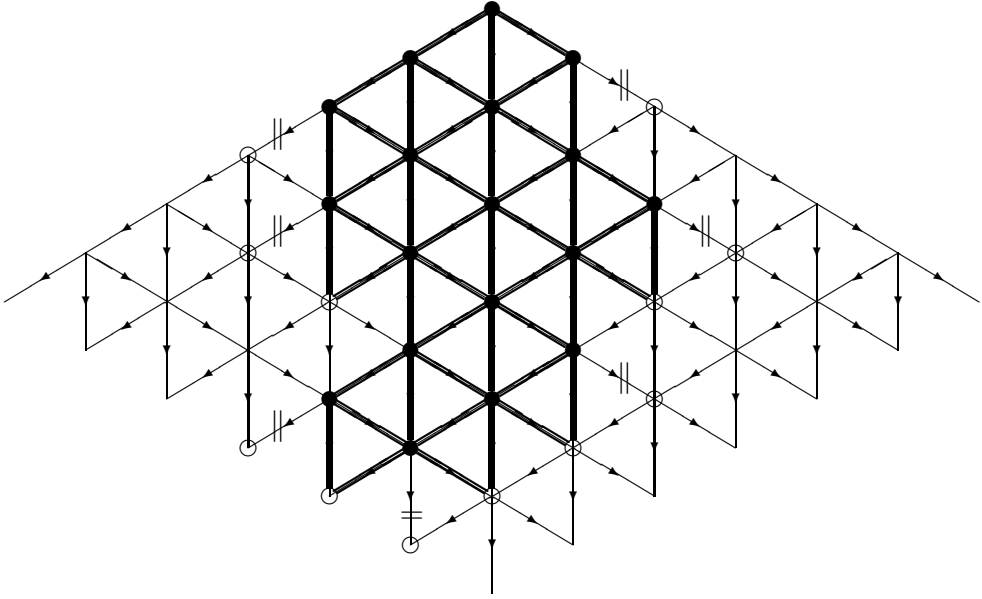

{}From the emperical extrapolation formulas it it clear that the
site-bond case is very similar to the site case and only a few
generalisations are necessary. Again we look at compact animals
and the ones we shall call {\em site-compact}
have the minimal perimeter. A site-compact animal is one in
which, as before, all occupied sites and bonds in a row are
consecutive and in addition {\em all possible bonds to sites
with more than one incident edge are present}.
Figure~\ref{fig-tsbani} shows such an animal. Clearly the
perimeter of such an animal is equal to the perimeter
of the identical {\em site} animal. Thus the animals with
minimal perimeter have $w_l =1$ (or $w_r =1$). Such animals
consist of consecutive occupied sites down the left-hand side
with most of the bonds emanating from these sites present. A few
of the bonds can be either present or absent, namely, the bond
from the top site pointing South-East and the bonds from the
last site pointing South-West or South, though in this latter
case at least one of the bonds has to be present. So all in all
there are three possible bond configurations from the last site
and two from the top site for a total of six possibilities.
Taking into account the animals with $w_r = 1$ we have proved

$$
d_{N,0} = 12
$$

{\large \bf The bond case}

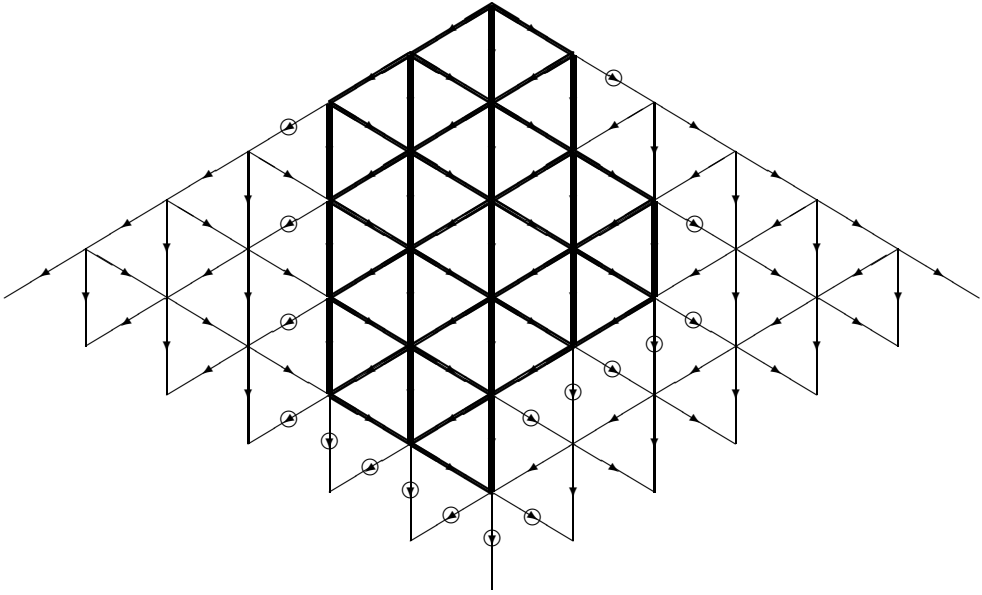
\begin{figure}[tb]
\begin{picture}(420,260)
\newsavebox{\swbb}
\newsavebox{\sebb}
\newsavebox{\ssbb}

\savebox{\swbb}(30,18){\thicklines \put(0,0.4){\line(5,3){30}}
\put(0,-0.8){\line(5,3){30}}}
\savebox{\sebb}(30,18){\thicklines \put(0,0.4){\line(-5,3){30}}
\put(0,-0.8){\line(-5,3){30}}}
\savebox{\ssbb}(4,36){\thicklines \put(-0.7,0){\line(0,1){35}}
\put(0.7,0){\line(0,1){35}}}

\multiput(240,42)(30,18){6}{\line(-5,3){180}}
\multiput(180,42)(-30,18){6}{\line(5,3){180}}
\multiput(210,24)(-30,18){6}
{\multiput(0,0)(30,18){6}{\line(0,1){36}}}

\multiput(60,132)(30,-18){6}
{\multiput(0,0)(30,18){6}{\vector(0,-1){2}}}
\multiput(45,141)(30,-18){6}
{\multiput(0,0)(30,18){6}{\vector(-3,-2){2}}}
\multiput(75,141)(30,-18){6}
{\multiput(0,0)(30,18){6}{\vector(3,-2){2}}}

\multiput(150,204)(30,18){2}{\usebox{\swbb}}
\multiput(150,168)(30,18){3}{\usebox{\swbb}}
\multiput(150,132)(30,18){3}{\usebox{\swbb}}
\multiput(150,96)(30,18){4}{\usebox{\swbb}}
\multiput(180,78)(30,18){3}{\usebox{\swbb}}
\put(240,222){\usebox{\sebb}}
\multiput(270,168)(-30,18){3}{\usebox{\sebb}}
\multiput(270,132)(-30,18){4}{\usebox{\sebb}}
\multiput(240,114)(-30,18){3}{\usebox{\sebb}}
\multiput(210,96)(-30,18){2}{\usebox{\sebb}}
\multiput(210,60)(-30,18){2}{\usebox{\sebb}}
\multiput(150,96)(0,36){3}{\usebox{\ssbb}}
\multiput(180,78)(0,36){4}{\usebox{\ssbb}}
\multiput(210,60)(0,36){5}{\usebox{\ssbb}}
\multiput(240,114)(0,36){3}{\usebox{\ssbb}}
\put(270,132){\usebox{\ssbb}}

\multiput(135,87)(0,36){4}{\circle{6}}
\multiput(150,79)(30,-18){3}{\circle{6}}
\multiput(165,69.5)(30,-18){2}{\circle{6}}
\put(225,51){\circle{6}}
\multiput(224.5,87.5)(30,18){3}{\circle{6}}
\multiput(240,97)(30,18){2}{\circle{6}}
\put(285,159){\circle{6}}
\put(255,213){\circle{6}}
\end{picture}
\caption{ \label{fig-tbani} \sf
A compact directed bond animal (thick
bonds) on the triangular
lattice with perimeter bonds marked by open circles.
}
\end{figure}

The first correction term for the bond case, $d_{N,0} = 2C_N-1$,
involve the Catalan numbers $C_N$ which equal the first correction
term for the square bond problem (Baxter and Guttmann 1988).
Bousquet-M\'{e}lou (1995) proved this result by noting that the
square bond correction term arise from compact bond animals of
directed height $N$. The first correction term for the triangular
bond problem can be found by generalising the arguments from
the square bond case. The first correction arise from compact
animals constructed as follows. Choose two paths $\omega_1$ and
$\omega_2$ consisting
of bonds pointing only South and South-West starting from the
origin and terminating at the same point on level $N$. The animal
obtained
by filling in all bonds between $\omega_1$ and $\omega_2$ has
height $N$ and perimeter $2N+1$. These animals are just the
{\em staircase animals} which are enumerated by the Catalan numbers
and give rise to the first square bond correction term. Obviously
the set of animals bounded by paths consisting
of South and South-East bonds also contribute to the first
correction term. The animal consisting entirely of south bonds (a line
of bonds down the center of the lattice) is the only animal included
in both sets. The first correction term is exactly due to
these $2C_N-1$ `staircase animals' on the triangular lattice.

\end{document}